\documentclass[fp,twocolumn]{jpsj3}
\usepackage{txfonts}
\usepackage{color,graphicx}
\usepackage{dcolumn}
\usepackage{amsmath}
\usepackage{amssymb}
\usepackage{bm}
\usepackage[normalem]{ulem}
\usepackage{chngcntr}
\usepackage{gensymb}
\newcommand{\NF}{NbFe$_{2}$}
\newcommand{\NFy}{Nb$_{1-y}$Fe$_{2+y}$}
\newcommand{\TF}{TaFe$_{2}$}
\newcommand{\TV}{TaV$_{2}$} 
\newcommand{\TFV}{Ta(Fe$_{1-x}$V$_{x})_{2}$}
\newcommand{\TFA}{Ta(Fe$_{1-x}$Al$_{x})_{2}$}

\newcommand{\TN}{$T_{\textrm{N}}$}
\newcommand{\muB}{$\mu_{\textrm{B}}$} 
\title{Quantum phase transitions and multicriticality in \TFV}
\author{\name{Manuel \surname{Brando}$^{1,}$\thanks{E-mail address: brando@cpfs.mpg.de},
\name{Alexander \surname{Kerkau}}$^{1}$,
\name{Adriana \surname{Todorova}}$^{1}$,
\name{Yoshihiro \surname{Yamada}}$^{2}$,
\name{Panchanan} \surname{Khuntia}$^{1}$,
\name{Tobias} \surname{F\"orster}$^{3}$,
\name{Ulrich} \surname{Burkhardt}$^{1}$,
\name{Michael} \surname{Baenitz}}$^{1}$
and \name{Guido} \surname{Kreiner}$^{1}$}
\inst{\address{$^{1}$Max Planck Institute for Chemical Physics of Solids, N\"othnitzer Stra\ss e 40, 01187 Dresden, Germany}\\
$^{2}$Division of Electronic Materials and Devices, Univesity of Hyogo, Shosha, Himeji 671-2280, Japan\\
$^{3}$Hochfeld-Magnetlabor Dresden (HLD-EMFL), Helmholtz-Zentrum Dresden-Rossendorf, 01314 Dresden, Germany}
\abst{We present a comprehensive study of synthesis, structure analysis, transport and thermodynamic properties of the C14 Laves phase \TFV. Our measurements confirm the appearance of spin-density wave (SDW) order within a dome-like region of the $x-T$ phase diagram with vanadium content $0.02 < x < 0.3$. Our results indicate that on approaching \TF\ from the vanadium-rich side, ferromagnetic (FM) correlations increase faster than the antiferromagnetic  (AFM) ones. This results in an exchange-enhanced susceptibility and in the suppression of the SDW transition temperature for $x < 0.13$ forming the dome-like shape of the phase diagram. This effect is strictly related to a significant lattice distortion of the crystal structure manifested in the $c/a$ ratio. At $x = 0.02$ both FM and AFM energy scales have similar strength and the system remains paramagnetic down to 2\,K with an extremely large  Stoner enhancement factor of about 400. Here, spin fluctuations dominate the temperature dependence of the resistivity $\rho \propto T^{3/2}$ and of the specific heat $C/T \propto -\log(T)$ which deviate from their conventional Fermi liquid forms, inferring the presence of a quantum critical point of dual nature.}
\kword{\TF, \NF, Laves phase, itinerant magnetism, strongly correlated metals, quantum criticality}
\begin{document}
\maketitle
\section{Introduction}
Quantum phase transitions (QPTs) in itinerant magnets have regained considerable attention after the discovery of unconventional superconductivity on the border of a spin density wave (SDW) order in the iron pnictides and chalcogenides~\cite{Sachdev1999,Stewart2011}. Similar to high-$T_{\textrm{c}}$ cuprates~\cite{Broun2008} and to heavy-fermion systems~\cite{Mathur1998,Yuan2003}, the superconductivity dome is found in the pressure-temperature $(p - T)$ phase diagram near the quantum critical point (QCP) at which the magnetic order is continuously suppressed to zero by pressure (or chemical substitution). Outside the FeAs and CuO families, itinerant antiferromagnetism, or SDW order, is rare in transition-metal compounds. The most studied example is chromium and its alloy series with vanadium Cr$_{1-x}$V$_{x}$, in which signatures of Fermi liquid (FL)\cite{Baym2004} breakdown have been observed at the QCP~\cite{Yeh2002}. On the other hand, QPTs and the associated quantum critical behavior have been explored in detail on nearly or weakly ferromagnetic (FM) materials like MnSi~\cite{Pfleiderer2007}, Ni$_3$Al/Ni$_3$Ga~\cite{Niklowitz2005},  ZrZn$_2$~\cite{Smith2008} and in the layered oxides such as the high-$T_{\textrm{c}}$ cuprates or in the ruthenates~\cite{Grigera2004}.

Promising candidates for the investigation of SDW QPTs are the Laves phases with hexagonal C14 crystal structure, in which an antiferromagnetic (AFM) ground state has been reported. Some examples are \NFy~\cite{Yamada1988,Crook1995}, \TFA~\cite{Yamada1990} or \TFV~\cite{Horie2010}. The system \NFy\ is of particular interest and it has been investigated in detail during the last years~\cite{Moroni2009,Duncan2010,Rauch2015}: It exhibits three magnetically ordered low-$T$ states within a narrow composition range $(-0.02 \leq y \leq 0.02)$ at ambient pressure. At slightly off-stoichiometric compositions, both towards the Fe-rich and the Nb-rich side, it has been reported to be FM and at $y \approx 0$ it assumes SDW order below 10\,K. The QCP is located at $y \approx -0.01$ at which non-FL (NFL) temperature dependencies of the heat capacity $(C/T \propto -\log T)$ and of the electrical resistivity $\rho \propto T^{3/2}$ were found, but no superconductivity~\cite{Brando2008}.

The system presented in this work, \TFV\, is very similar to \NFy\ and also shows peculiar magnetic properties: In the range $0.05 \leq x \leq 0.25$ it was reported to be antiferromagnetic with a peculiar dome-like shape of the AFM phase in the $x - T$ phase diagram. For $x < 0.05$ the magnetic susceptibility of the system, measured with a magnetic field of 1\,T, did not show any phase transition but very high values for a band magnet, indicating the proximity of \TF\ to a FM instability~\cite{Horie2010}. \TF\ was therefore considered to be paramagnetic (PM) with strong AFM and FM spin fluctuations, similarly to \NF\ which, despite the SDW order, shows a very high value of the magnetic susceptibility $\chi \approx 0.02$. This corresponds to a Stoner enhancement factor of 180 and a Wilson ratio of 60, and explains the vicinity of \NF\ to FM phases. The similarity between \TF\ and \NF\ is simply suggested by the homologous chemical and electronic structures, but since the atomic volume of \TF\ is about 13.23\,\AA$^{3}$ and thus slightly smaller than that of \NF\ (13.35\,\AA$^{3}$) we expect the properties of \TF\ to be a little different from those of \NF. 

In this article, we confirm that the system \TFV\ displays SDW order for $0.05 \leq x \leq 0.25$ and show that the dome-like shape of the SDW phase is due to the competition between AFM and FM exchange interactions. In particular, below $x = 0.15$, FM correlations start to increase rapidly and suppress the transition temperature \TN\ down to zero at $x \approx 0.02$ where the system is PM with a very high susceptibility $\chi > 0.02$. At this V content we also observe the same NFL behavior seen in \NF\ at the QCP, i.e. $\rho \propto T^{3/2}$ and $C/T \propto -\log(T)$, indicating that the QCP at $x \approx 0.02$ in \TFV\ and the QCP in \NF\ have the same origin and nature. The fact that in both systems the QCP is located where both the AFM and the FM energy scales vanish suggest that the QCP has an intriguing dual nature and is therefore a multicritical point.
\section{Experimental details}
\subsection{Sample preparation}
A series of alloys with the general composition \TFV\ with 0 $\leq$ \emph{x} $\leq$ 0.35 (see Table~\ref{table}) was prepared from Ta, Fe foil, (both Alfa
Aesar, Puratronic 99.9995\,\%) and vanadium foil (Alfa Aesar, Puratronic, 99.8\,\%) by argon arc-melting on a water-cooled copper hearth (Centorr Series 5BJ, Centorr Vacuum Industries). Prior to the preparation the argon atmosphere inside the arc-melter has been purified by melting an ingot of titanium several times in an adjacent recess of the copper hearth. The specimens of 2\,g total mass were melted several times to ensure compositional homogeneity. The weight loss after arc-melting was smaller than 0.5\,wt.\%. For the following heat treatment each specimen was encapsulated in a weld-sealed tantalum ampoule (Plansee AG). The Ta ampoules in turn were jacketed by a fused silica tube under a pressure of about 250\,mbar pure argon, and annealed isothermally at 1150\,$^{\circ}$\,C for 30 days. After annealing the samples were quenched in water by shattering the fused silica ampoules.
\subsection{Alloy characterization}
Atomic emission spectrometry with plasma excitation (ICP-OES, Varian, Vista RL) was used to verify the bulk composition of the alloys after heat treatment. For each analysis 20\,mg of the material were dissolved in a mixture of hydrofluoric and nitric acid.

All samples were characterized by powder X-ray diffraction (PXRD) using an image-plate Guinier camera (Huber G670, CoK$\alpha_1$ with $\lambda$ = 1.78892\,{\AA}). Parts of the samples were ground in an agate mortar and homogeneously dispersed and loaded between two n-hexane/vaseline coated mylar foils. The powder diffraction intensities, as well as the peak positions were obtained using WinXpow routines~\cite{Stoe}. Indexing and refinement of the lattice parameter by a least-squares refinement of the diffraction angles in the range $20^{\circ} < 2\theta < 100^{\circ}$ were done after calibration with Si ($a$ = 5.43119(1)\,{\AA}) or LaB$_6$ ($a$ = 4.15692(1)\,{\AA}) as an internal standard. Vanadium rich samples ($x_{nom.}$ = 0.3 and 0.35) were measured in addition at the high resolution powder diffraction beamline ID31 at the European Synchrotron Radiation Facility (ESRF) in Grenoble/France. The X-ray wavelength was refined to $\lambda$ = 0.4007(3)\,{\AA} using LaB$_6$ as reference material. For Rietveld refinements $2\theta$-scans with $4^{\circ} < 2\theta < 58^{\circ}$ were performed. Structure refinement was done by performing full-matrix least-square refinements with the aid of the program Jana2006~\cite{Jana2006}.

Metallographic analysis was performed for the annealed specimens. Specimens of about 3\,mm diameter were embedded in conducting bakelite resin (Polyfast, Struers) and then mechanically grinded by abrasive papers (SiC, 500, 1000 and 2000 grit) using water as lubricant. Polishing was done using a slurry of 9, 6 and 3\,$\mu$m diamond paste suspended in water and a mixture of 80\,\% ethyl alcohol, 10\,\% propyl alcohol and 10\,\% ethylene glycol as lubricant. The final polishing step was performed with colloidal silica (0.05\,$\mu$m) suspended in water. The microstructure was analyzed using an optical microscope (Zeiss Axioplan 2) with brightfield, polarized light and differential interference contrast (DIC) and a scanning electron microscope (Philips XL30) equipped with an energy-dispersive X-ray spectrometer (EDAX) to check for homogeneity.
\subsection{Experimental setup}
The resistivity, obtained by a standard four-terminal ac technique, and the specific heat were measured in a Physical Property Measurement System (PPMS) by Quantum Design. We used a Vibrating Sample Magnetometer (VSM) SQUID to measure the temperature and field dependence of the magnetization.
\section{Results}
\subsection{Phase analysis}
\TV\ and \TF\ are, according to their binary phase diagrams, both Laves phases~\cite{Guzei1970,Kuo1953,Pauling2002} at 1150\,$^{\circ}$\,C. \TV\ crystallizes cubic in the C15 structure type, while \TF\ adopts the hexagonal C14 polytype. Crystallographic data of \TF\ has been reported several times~\cite{Pauling2002}. The lattice parameters \emph{a} and $\emph{c}$ of \TF\ are compared for convenience in Table~S1 and Fig.~S1 in the Supplemental Materials. There, it is shown that the lattice parameters reported in the literature vary considerable. The lattice parameter \emph{a} ranges from 4.80\,{\AA} to 4.86\,{\AA} and \emph{c} from 7.83\,{\AA} to 7.91\,{\AA} yielding an average unit cell volume with a large standard deviation of $V =
(158.7\pm 1.6) \,{\mathrm{\AA}}^3$. Horie~\textit{et al.} have shown that, if TaFe$_{2}$ is alloyed with small amounts of V, it crystallizes at least up to \textit{x} = 0.35 with \emph{x} in \TFV\ in the C14 structure type~\cite{Horie2010}. Otherwise nothing is known about the pseudobinary section \TV-\TF, no ternary compounds and no phase diagram have been reported.

The results of the phase analysis for an alloy series \TFV\ with $0\leq\ x \leq 0.35$ as obtained from X-ray powder diffraction, metallographic examinations and quantitative chemical analysis by ICP-OES are summarized in
Table~\ref{table} together with the transition temperatures and residual resistivity ratios (RRRs). The difference in at.\,\% between the nominal and the experimental content for each sample is shown for Ta, Fe and V  in figures S2, S3, and S4, respectively, in the Supplemental Materials. The nominal compositions for all samples agree well with the results of the analytical
investigation. The substitution parameter $x$ in \TFV\ obtained from the chemical analysis as given in column 3 of Table~\ref{table} is used in the following discussion.

The present study confirms the work of Horie \textit{et al.}, i.e., the formation of a C14 Laves phase in the range $0 \leq\ x \leq 0.35$. However,
together with the expected hexagonal C14 Laves phase, powder X-ray diffraction patterns, BSE images and EDXS measurements show evidence of traces of a second phase ($< 0.1\,\%$) with Ti$_{2}$Ni structure type (space group $Fd\bar{3}m$, $a \approx 11.24\,{\mathrm{\AA}}$) in several samples. This phase was found in the Co-Cr-Nb system~\cite{Kerkau2012} and is already known from our investigation on the series \NFy\ (see Ref.~\citeonline{Moroni2009}). A detailed analysis of this impurity phase revealed that this phase is richer in Ta
compared to the Laves phase. In all samples containing this phase we have observed a small signature of a FM phase transition around 80\,K. Moreover, this signature is stronger in samples cut from the upper/lower edge of the pellet instead of from the middle zone. Similar effects were observed and characterized in \NFy~\cite{Moroni2009}. Therefore, we used for our thermodynamic and transport experiments only samples from the middle of the pellets where this FM phase was found to be negligible.
\begin{figure}[t]
	\begin{center}
		\includegraphics[width=0.95\columnwidth]{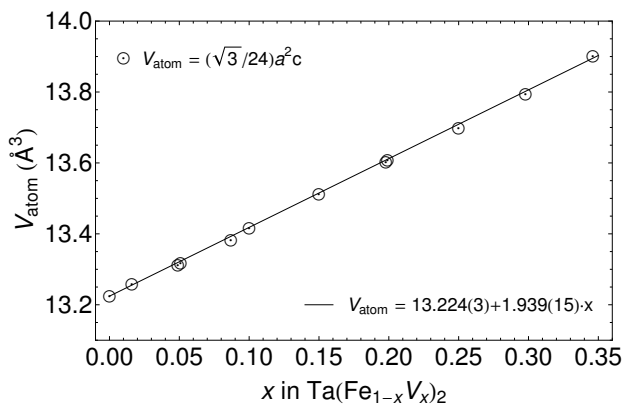}
	\end{center}
	\caption{Mean atomic volume per atom of the hexagonal C14 Laves phase \TFV\ calculated from the unit cell volumes displayed in Tab.~\ref{table} versus the experimental substitution parameter $0 \leq x \leq 0.35$. The line is a linear fit according to Vegard's volume rule as given in Eq.~\ref{eq:volume}.}
	\label{meanatomicvolume}
\end{figure}

The arithmetic mean for the lattice parameters \emph{a} and \emph{c}, the unit cell volume and the mean atomic volume for \TF\ samples of the present work (Nos. 1 to 7 in Table~\ref{table}) are \emph{a} = 4.8247(9)\,{\AA}, \emph{c} = 7.875(2)\,{\AA}, \emph{V} = 158.75(5)\,{\AA}$^{3}$ and $V_{\mathrm{atom}} = 13.229(4)\,{\mathrm{\AA}}^3$. These values compare well with the average values taken from the literature except that the data of this work are much more precise. The mean atomic volume $V_{\mathrm{atom}}$ for samples Nos. 8 to 18 and the average value obtained from samples Nos. 1 to 7 are plotted versus the substitutional parameter \emph{x} in Fig.~\ref{meanatomicvolume}. The mean atomic volume $V_{\mathrm{atom}}$ increases with increasing \emph{V} content as expected from the atomic radii of Fe (1.26\,{\AA}) and \emph{V} (1.35\,{\AA}),~\cite{Pauling1927} and behaves according to Vegard's volume rule. Fitting the points for all \emph{x} except \emph{x} = 0 gives for \TF\ an extrapolated mean atomic volume of 13.223(3)\,{\AA}$^{3}$, in excellent agreement with the value obtained from the vanadium free samples. A fit using all data yields:
\begin{equation}
V_{\mathrm{atom}}({\mathrm{\AA}}^3)=13.224(3) + 1.939(15) \cdot x.
\label{eq:volume}
\end{equation}

\begin{table*}[t]
	\caption{Results of the phase analysis for all \TF\ and \TFV\ samples: nominal \emph{x}, ICP-OES results, lattice parameter \emph{a} and \emph{c}, unit cell volume V, Neel temperature $T_{N}$, residual resistivity data RRR. The column labelled 'FM i.p.' indicates whether weak signatures of a FM impurity phase at about 80\,K were observed or not.}
	\label{table}
	\begin{center}
		\begin{tabular}{llllllllll}
			\hline
			\\
			No. & $x_{nom.}$ & ICP-OES & $a$ (\AA) & $c$ (\AA) & V (\AA$^{3}$) & $T_{N}$(K) & RRR & FM i.p.\\ 
			\hline
			1  & 0 & Ta$_{1.011\pm0.025}$Fe$_{1.989\pm0.025}$ &  4.8242(5) & 7.8739(6) & 158.70(3) & FM (20\,K) & -- & yes\\ 
			2  & 0 & Ta$_{0.990\pm0.011}$Fe$_{2.010\pm0.011}$ &  4.8246(3) & 7.8747(3) & 158.74(2) & FM (15\,K) & 3.70 & yes\\ 
			3  & 0 & Ta$_{1.008\pm0.014}$Fe$_{1.992\pm0.014}$ &  4.8237(2) & 7.8732(3) & 158.65(2) & 0 & 4.65 & yes\\ 
			4  & 0 & Ta$_{0.987\pm0.006}$Fe$_{2.013\pm0.006}$ &  4.8245(3) & 7.8748(5) & 158.73(2) & 8 & -- & yes\\ 
			5  & 0 & Ta$_{0.993\pm0.011}$Fe$_{2.007\pm0.011}$ &  4.8241(8) & 7.8747(10) & 158.71(5) & 10 & 3.80 & yes\\ 
			6  & 0 & Ta$_{0.993\pm0.010}$Fe$_{2.007\pm0.010}$ &  4.826(1) & 7.880(3) & 158.92(11) & FM (70\,K) & -- & yes\\ 
			7  & 0 & Ta$_{1.017\pm0.017}$Fe$_{1.983\pm0.017}$ &  4.8257(6) & 7.8762(9) & 158.85(4) & 5 & -- & yes\\ 
			\hline
			8  & 0.02 & Ta$_{1.008\pm0.022}$(Fe$_{0.978\pm0.012}$,V$_{0.016\pm0.001}$)$_{2}$ &  4.8293(4) & 7.8796(5) & 159.15(3) & 0 & 1.93 & tiny\\ 
			9  & 0.05 & Ta$_{1.023\pm0.027}$(Fe$_{0.939\pm0.015}$,V$_{0.049\pm0.001}$)$_{2}$ &  4.8366(4) & 7.8871(5) & 159.78(3) & 30 & -- & tiny\\ 
			10 & 0.05 & Ta$_{1.017\pm0.021}$(Fe$_{0.940\pm0.011}$,V$_{0.051\pm0.002}$)$_{2}$ &  4.8374(5) & 7.8885(5) & 159.86(3) & 32 & 1.45 & yes\\ 
			11 & 0.09 & Ta$_{1.002\pm0.018}$(Fe$_{0.912\pm0.010}$,V$_{0.087\pm0.002}$)$_{2}$ &  4.8459(4) & 7.8991(6) & 160.64(3) & 65 & 1.27 & no\\ 
			12 & 0.10 & Ta$_{1.017\pm0.021}$(Fe$_{0.891\pm0.012}$,V$_{0.100\pm0.003}$)$_{2}$ &  4.8497(4) & 7.9058(5) & 161.03(2) & 68 & 1.20 & yes\\ 
			13 & 0.15 & Ta$_{1.020\pm0.005}$(Fe$_{0.841\pm0.003}$,V$_{0.150\pm0.001}$)$_{2}$ &  4.8615(4) & 7.9242(5) & 162.19(3) & 64 & 1.15 & yes\\ 
			14 & 0.20 & Ta$_{1.002\pm0.008}$(Fe$_{0.799\pm0.006}$,V$_{0.199\pm0.004}$)$_{2}$ &  4.8726(5) & 7.9439(8) & 163.34(4) & 42 & -- & no\\ 
			15 & 0.20 & Ta$_{0.999\pm0.017}$(Fe$_{0.801\pm0.011}$,V$_{0.198\pm0.005}$)$_{2}$ &  4.8716(5) & 7.9432(7) & 163.26(3) & -- & 1.10 & --\\ 
			16 & 0.25 & Ta$_{1.002\pm0.016}$(Fe$_{0.748\pm0.010}$,V$_{0.250\pm0.006}$)$_{2}$ &  4.8826(3) & 7.9644(5) & 164.44(2) & 17 & 1.09 & no\\ 
			17 & 0.30 & Ta$_{1.002\pm0.024}$(Fe$_{0.700\pm0.016}$,V$_{0.298\pm0.010}$)$_{2}$ &  4.8930(3) & 7.9856(5) & 165.57(2) & 0 & -- & no\\ 
			18 & 0.35 & Ta$_{1.011\pm0.018}$(Fe$_{0.648\pm0.012}$,V$_{0.346\pm0.008}$)$_{2}$ &  4.9049(2) & 8.0093(3) & 166.87(2) & 0 & 1.09 & no\\ 
			\hline
		\end{tabular}
	\end{center}
\end{table*}

\begin{figure}[!ht]
	\begin{center}
		\includegraphics[width=0.95\columnwidth]{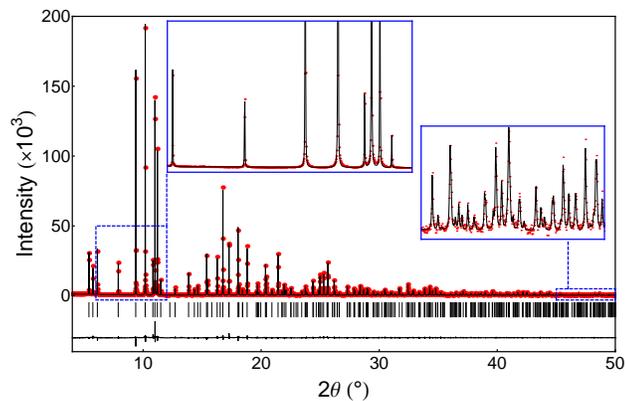}
	\end{center}
	\caption{Synchrotron powder X-ray diffraction pattern (red points) and fit from the Rietveld refinement (solid line), Bragg reflection marker and difference curve of C14 Ta(Fe$_{0.7}$V$_{0.3}$)$_{2}$ (sample No. 17). Two insets show zoomed parts of the pattern at low and high	$2\theta$, respectively.}
	\label{Plot_Refinement_AK160}
\end{figure}

The crystal structure of C14 \TFV\ in dependence of the substitution parameter \emph{x} has been studied by means of Rietveld analysis to obtain information about deviations from an idealized crystal structure $AB_{2}$ of C14 type. In the latter case all interatomic distances $d(B-B)$ and all distances $d(A-A)$ are equal, respectively. In addition, the ratio $(c/a)_{ideal}$ is $\surd(8/3)\approx 1.633$. Crystallographic data, fractional atomic coordinates, isotropic displacement parameters and selected interatomic distances for \TF\ (sample No. 3) and Ta(Fe$_{0.7}$,V$_{0.3}$)$_{2}$ (sample No. 17) are given in Tables S5, S6 and S7 in the Supplemental Materials. The powder X-ray diffraction pattern for C14 Ta(Fe$_{0.7}$V$_{0.3}$)$_{2}$ and the
fit from the Rietveld refinement are shown in Fig.~\ref{Plot_Refinement_AK160}.
\begin{figure}[b]
	\begin{center}
		\includegraphics[width=0.6\columnwidth]{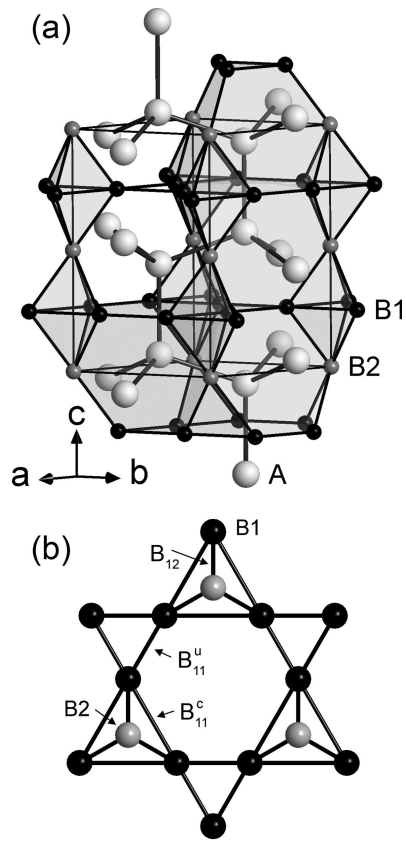}
	\end{center}
	\caption{a) Crystal structure of a C14 Laves phase $AB_{2}$. b) Partial structure of \emph{B} atoms with $\emph{x}(B1) > 1/6$ in the structure of C14	$AB_{2}$ indicating the different B-B distances between \emph{B}1 at 6\emph{h} and \emph{B}2 at 2\emph{a}.}
	\label{C14_Structure_Net}
\end{figure}

The crystal structure of a C14 Laves phase with the general composition $AB_{2}$ for a non-ideal case is shown in Fig.~\ref{C14_Structure_Net}. The smaller \emph{B} atoms occupy the vertices of a six-connected network composed of $B_{4}$ tetrahedra, which are joined alternately point-to-point and base-to-base, thereby forming infinite chains along \emph{c} of apically-fused
trigonal bipyramids. These chains are linked together in the \emph{a},\emph{b} plane, thus forming large truncated $B_{12}$ tetrahedra. The larger \emph{A} atoms occupy the center of these polyhedra and form a four-connected net of wurzite type. The \emph{A} and the \emph{B} atoms are coordinated by Friauf polyhedra (CN = 16) and by icosahedra (CN = 12), respectively. The crystal structure comprises triangular and Kagom\'e{} layers, which are stacked along \emph{c}. The Kagom\'e{} layers are composed of the smaller \emph{B} atoms, which are located at the Wyckoff site 6\emph{h} (\emph{B1} at (\emph{x},\,2\emph{x},\,1/4) with $x_{ideal} = 1/6$). One set of triangular layers is formed by the remaining \emph{B} atoms at the Wyckoff site 2\emph{a} (\emph{B2} at (0,\,0,\,0)). The other set is formed by \emph{A} atoms at the
Wyckoff site 4\emph{f} (\emph{A} at (1/3,\,2/3,\,\emph{z}) with $z_{ideal}=9/16$).

Rietveld analysis for the C14 phase with different substitution parameter \emph{x} show that the \emph{A} sites are exclusively occupied by the large Ta atoms ($r_{Ta} = 1.47\,{\AA}$)~\cite{Pauling1927} and the \emph{B} sites by the smaller Fe and V atoms. Synchrotron powder X-ray data for the V-rich samples with $x = 0.3$ and $0.35$ do not indicate significant preferential site occupation of Fe or V at the 6\emph{h} or 2\emph{a} sites. Although, high angle data up to $sin(\theta)/\lambda = 1.21{\mathrm{\AA}}^{-1}$ have been used for the refinement of the site occupation factors of Fe and V at the \emph{B} sites 6\emph{h} and 2\emph{a}, a small preferential site occupation of less than 10\,\% deviation from fully random mixed occupation of the 2\emph{a} and 6\emph{h} sites cannot be excluded. Owing to the similar X-ray atomic form factors of Fe ($Z = 26$) and V ($Z = 23$) and the tiny difference between ordered and disordered models in the electron density distribution it is not possible to exclude completely partial disorder for the C14 phase with less V
than $x = 0.3$. A finite temperature approach based on first principles calculations, which is not discussed here, indicates a preference of a few percent for the minority component V to occupy preferentially the 2\emph{a} site. Because no experimental data are available to verify such a preferential site occupation and knowing from experiment and theory that the upper limit is about 10\,\% deviation from fully random occupation, the C14 phase is assumed in the following discussion to crystallize with fully random mixed occupation of the \emph{B}-sites by Fe and V atoms, when quenched from 1150\,$^{\circ}$\,C.

\begin{figure}[b]
	\begin{center}
		\includegraphics[width=0.95\columnwidth]{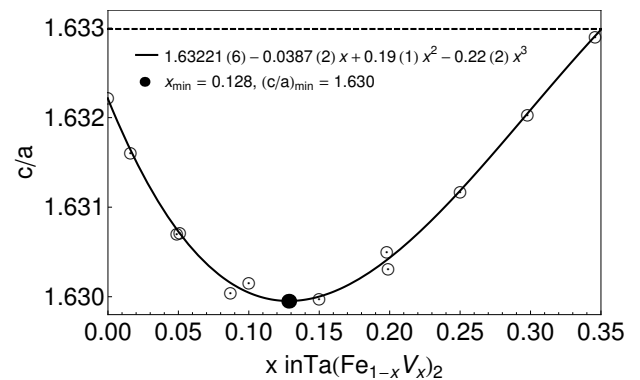}
	\end{center}
	\caption{\emph{c/a} ratio of C14 \TFV. The dashed line corresponds to $(c/a)_{ideal} \approx 1.633$.}
	\label{caratioexpfit}
\end{figure}

A significant lattice distortion of \TFV\ is manifested by the \emph{c/a} ratio, which is shown in Fig.~\ref{caratioexpfit}. The \emph{c/a} ratio for  TaFe$_{2}$ is slightly smaller than the ideal ratio 1.633. With increasing
substitutional \emph{x} the \emph{c/a} ratio decreases until a minimum at $x = 0.13$ and $c/a = 1.630$ is reached. Then the \emph{c/a} ratio increases until at $x = 0.35$ the ideal ratio is nearly reached. Contrary to the \emph{c/a} ratio, the fractional atomic parameter \emph{x}(Fe,V) and \emph{z}(Ta) are nearly independent from the composition. For \TF, $x(\textnormal{Fe,V})= 0.1696(1)$ and $z(\textnormal{Ta}) = 0.5646(1)$ and for Ta(Fe$_{0.7}$,V$_{0.3)}$)$_{2}$ $x(\textnormal{Fe,V})= 0.1714(7)$ and and $z(\textnormal{Ta}) = 0.5629(3)$ is found. However, $x(\textnormal{Fe,V})$ and
$z(\textnormal{Ta})$ deviate significantly from the ideal values $x(\textnormal{Fe,V}) = 1/6$ and $z(\textnormal{Ta}) = 9/16$ indicating a structural distortion.

\begin{figure}[t]
	\begin{center}
		\includegraphics[width=0.95\columnwidth]{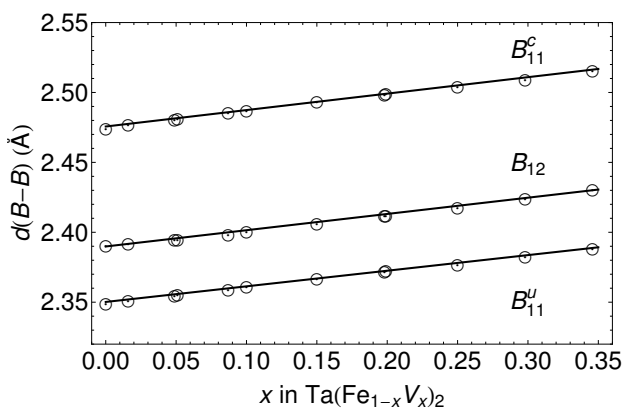}
	\end{center}
	\caption{Interatomic
		distances $B-B$ with \emph{B} = Fe and V in the crystal structure of
		C14 Ta(Fe$_{0.7}$V$_{0.3}$)$_{2}$ versus the experimental
		substitution parameter $0 \leq x \leq 0.35$.}
	\label{PlotDistancesBB}
\end{figure}

The interatomic distances between the smaller \emph{B} atoms in the Kagom\'e{} layer are labeled $B_{11}^u$ and $B_{11}^c$ in Fig.~\ref{C14_Structure_Net} and the distance between the atoms \emph{B} atoms in the Kagom\'e{} and the triangular layers $B_{12}$. Superscripts \emph{u} and \emph{c} denote uncapped and capped triangles in the Kagom\'e{} layers through the \emph{B}2 atoms in the triangular layers. The distances are then given by the following equations and plotted in Fig.~\ref{PlotDistancesBB} versus the substitution parameter \emph{x} using 0.171 for the fractional parameter \emph{x}:
\begin{eqnarray}
d(B_{11}^u) &=& (1-3x)a\\
d(B_{11}^c) &=& 3xa\\
d(B_{12}) &=& \sqrt{3(1/3-x)^2a^2+c^2/16}.
\end{eqnarray}
The interatomic distances $B_{11}^u < B_{12} < B_{11}^c$  are increasing with increasing V content. The deviation from the ideal partial structure of the (Fe,V) atoms can be expressed by $\Delta B_{11}$ and $\Delta B_{12}$ as given by the following equations:
\begin{equation}
\label{eq:b_distort} \Delta B_{11} = \frac{2(B_{11}^c -
	B_{11}^u)}{B_{11}^c + B_{11}^u} \textnormal{      and      } \Delta
B_{12} = \frac{2(B_{11}^c - B_{12})}{B_{11}^c + B_{12}}.
\end{equation}
$\Delta B_{11}$ describes the distortion of the Kagom\'e layer, which is about 5\,\% and does not depend on the composition, whereas the distortion along \emph{c} is given by $\Delta B_{12}$. The latter depends on the composition and is maximal 3.5\,\% at $x(\textnormal{V}) = 0.13$, i.e., at the minimum of the \emph{c/a} ratio. A plot of the relative distortion $\Delta B_{11}$ is shown in Fig. S6 in the Supplemental Materials.

In summary, the Laves phase C14 \TFV\ forms at least up to $x(\textnormal{V}) = 0.35$, shows random disorder of Fe and V at the icosahedral coordinated sites and deviates from an ideal C14 type structure with a maximal distortion at the composition $x(\textnormal{V}) = 0.13$.
\subsection{Resistivity}
\begin{figure}[ht!]
\begin{center}
\includegraphics[width=0.95\columnwidth]{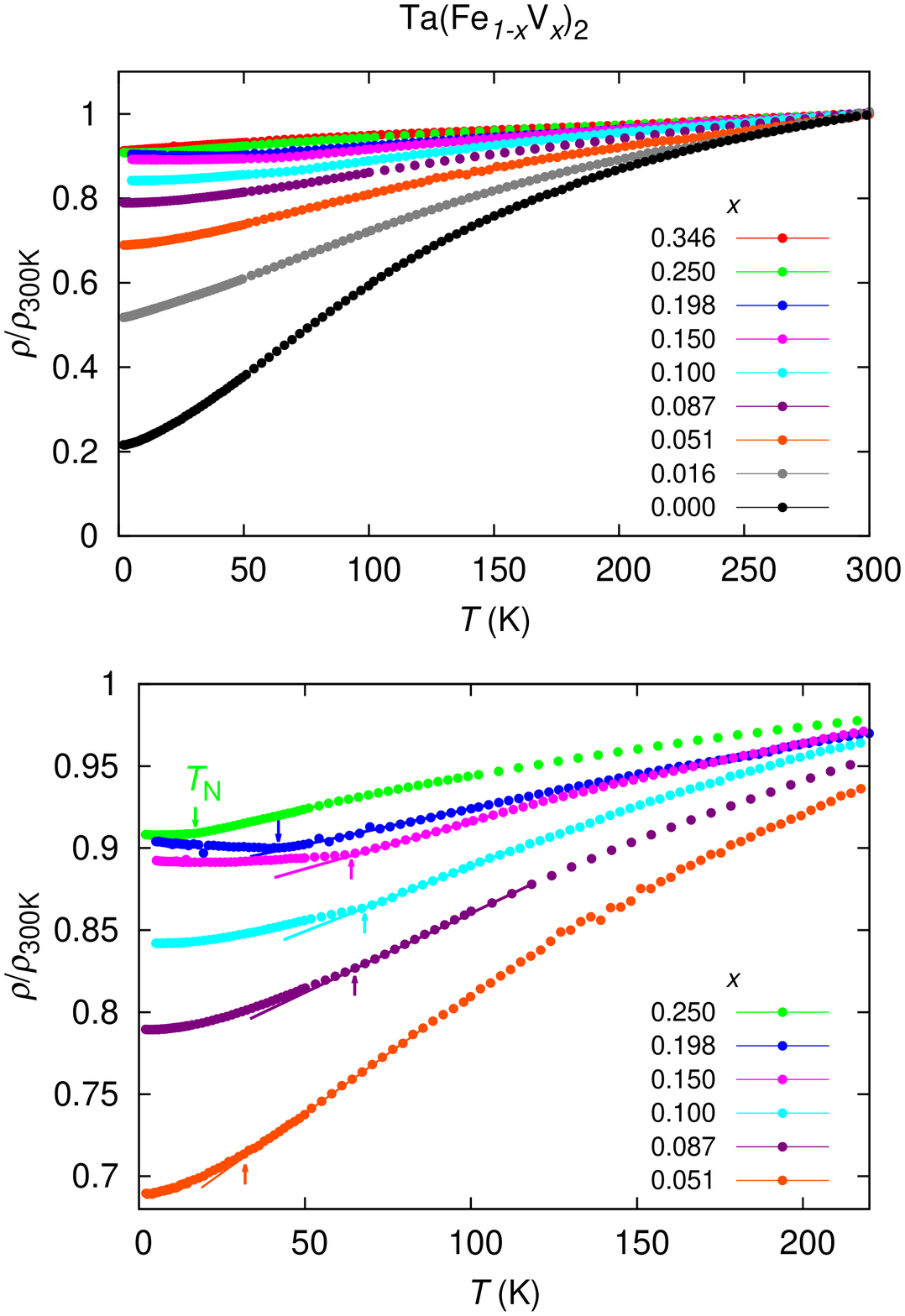}
\end{center}
\caption{Temperature dependence of the electrical resistivity $\rho(T)$ normalized to its value at 300\,K, $\rho_{\textrm{300K}}$, of several \TFV\ samples with $0 \leq x < 0.35$ for $0 \leq T \leq 300$\,K (upper panel) and $0 \leq T \leq 220$\,K (lower panel). The arrows indicate the position of the SDW transition temperature \TN\ at which $\rho(T)$ shows a kink. We have emphasized this behavior by extrapolating the experimental points just above \TN\ with a straight line.}
\label{resistivity}
\end{figure}
The temperature dependence of the resistivity $\rho(T)$ of several samples of \TFV\ with $x$ between 0 and 0.35 is shown in the upper panel of Fig.~\ref{resistivity}. The resistivity has been normalized to its value at 300\,K, $\rho_{\textrm{300K}}$. We have determined the absolute value of the resistivity for one sample of \TF\ at room temperature and this is $(100 \pm  5)$\,$\mu\Omega$cm. This value is the same found for \NF~\cite{Crook1995,Moroni2009} and yields a residual resistivity of about 20\,$\mu\Omega$cm. All samples show metallic behavior, the resistivity decreases with decreasing temperature. However, this decrease is smaller in samples with high $x$ and stronger in samples with lower $x$, because of the scattering due to disorder. From this curves we have evaluated the RRR which is displayed in Tab.~\ref{table}: As expected, the RRR exhibits a systematic decrease with increasing $x$. The RRR varies from 1.09 to a maximum of 4.65 for a sample of \TF\ with unit-cell volume very close to that expected for pure \TF\ (cf. Eq.~\ref{eq:volume}).

In the lower panel of Fig.~\ref{resistivity} we zoom into the low-$T$ region. The resistivity curves show clear kinks at which $\rho(T)$ deviates from its expected trend: in fact, $\rho(T)$ starts increasing with decreasing temperature suggesting the opening of a gap at the Fermi level, typical of SDW phase transitions. This is corroborated by the fact that the SDW transition temperatures \TN\ (indicated by arrows) match quite well those observed in our susceptibility experiments (shown later) and those published in literature~\cite{Horie2010}. This effect is similar in all samples, but it is better seen in samples with higher $x$ suggesting a correlation between the size of the SDW gap and $x$. In a mean-field approach the SDW gap is proportional to the transition temperature, but we can not find a direct correlation here~\cite{Gruener1994}: For instance, the samples with $x = 0.100$ and $x = 0.087$ have almost the same transition temperatures but the SDW signature in resistivity is stronger in the sample with $x = 0.100$ than in that with $x = 0.087$.
\subsection{Susceptibility and DOS}
\begin{figure}[b]
\begin{center}
\includegraphics[width=0.95\columnwidth]{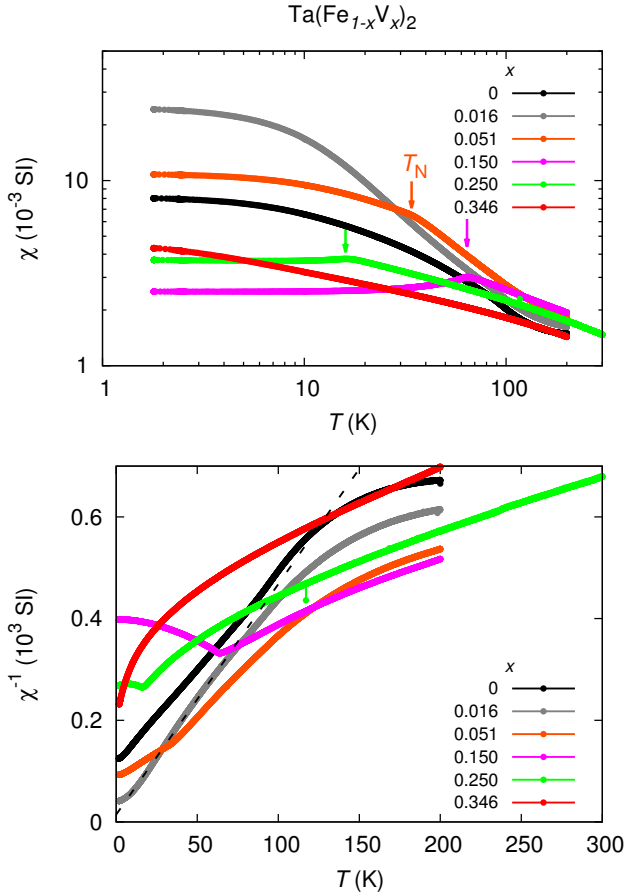}
\end{center}
\caption{Uniform susceptibility $\chi = M / B$ at $B = 1$\,T (upper panel) and $\chi^{-1}(T)$ (lower panel) plotted versus temperature for six samples with $0 \leq x \leq 0.346$. The arrows indicate the SDW transition temperature \TN. The dashed black line in the lower panel is a Curie-Weiss fit to the data for the sample with $x = 0.016$ below 70\,K.}
\label{susceptibility}
\end{figure}
\begin{figure}[t]
\begin{center}
\includegraphics[width=0.65\columnwidth, angle=-90]{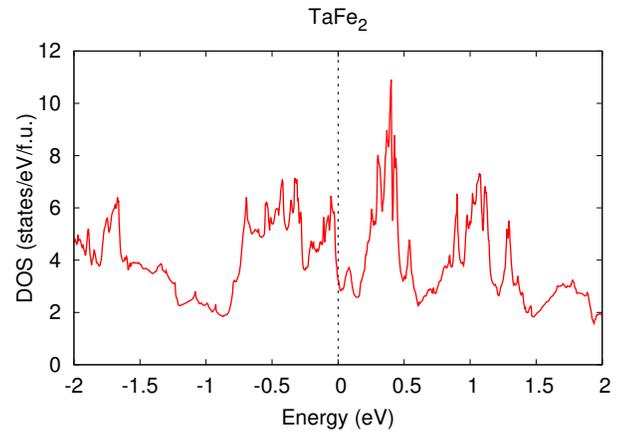}
\end{center}
\caption{Calculated density of states (DOS) of \TF.}
\label{dos}
\end{figure}

To get insight into the magnetic properties of \TFV\ we have measured the temperature and field ($B$) dependencies of the magnetization ($M$). The uniform susceptibility $\chi = M / B$ at $B = 1$\,T is plotted in Fig.~\ref{susceptibility} (upper panel) for selected samples and temperatures from 300 to 2\,K: the arrows mark the peak in $\chi$ at the N\'eel temperature \TN, indicating a phase transition from a PM state into a low-$T$ SDW state. These transition temperatures are drawn in the phase diagram of Fig.~\ref{phase_diagram}. At high and low V contents no phase transition is observed and the ground state is paramagnetic. However, there is a great difference between the values of the susceptibility for $x = 0.346$ (red points) and for $x = 0.016$ (grey points) at 2\,K. In the sample with $x = 0.016$ the susceptibility is much larger and is enhanced by spin fluctuations by a factor of about 400 (Stoner factor) compared to the bare susceptibility ($\approx 5\times 10^{-5}$\,SI) estimated from band structure calculations. In fact, we have calculated the band structure of \TF\ using the full-potential local-orbital FPLO code~\cite{Koepernik1999} on a $20\times 20\times 20$ $k$-mesh. The exchange and the correlation potentials were estimated using the local density approximation (LDA)~\cite{Perdew1992} with the structural data for the sample No. 3 in Tab.~\ref{table}. Our calculations are in good agreement with those by Diop \textit{et al.}~\cite{Diop2015}. The LDA gives 5 bands crossing the Fermi level. The density of state (DOS) for \TF\ is shown in Fig.~\ref{dos}. This is dominated by the iron $3d$ states at the Wyckoff site $6h$, i.e., within the Kagom\'e layers. At the Fermi energy it has a value of 3.3\,states/eV/f.u. which corresponds to a susceptibility of $5.61\times 10^{-5}$\,SI. The high value of the Stoner factor confirms that the system at $x \approx 0$ is close to a FM instability as it was observed in the homologous compound \NF\ in which a Stoner factor of about 180 was estimated~\cite{Brando2008}.

The stoichiometric sample No. 3 shows no phase transition down to 2\,K and a slightly smaller susceptibility than the one with $x = 0.016$ (black points in Fig.~\ref{susceptibility}). We did not observe any transition even at lower fields. This agrees well with the reported measurements of Ref.~\citeonline{Horie2010} and suggests that \TF\ is PM. However, measurements performed on all other samples with nominal vanadium content $x = 0$ indicate that the ground state is extremely sensitive to the composition: we could observe FM as well as AFM behavior in samples whose difference in composition $\Delta x$ is just 0.003 (cf. samples No. 2 and No. 5). For this reason, it is difficult to conclude from our investigations what is the real ground state of \TF.

The magnetic susceptibility for samples close to $x = 0$ are dominated mainly by spin fluctuations. According to the self-consistent renormalization theory~\cite{Moriya1985} for nearly and weakly FM metals the susceptibility follows a Curie-Weiss law which does not originate from local moments but from the coupling between different modes of spin fluctuations (cf. Eq.~\ref{A}). To know how large the fluctuating moment is in the sample with $x = 0.016$, we have plotted $\chi^{-1}$ vs $T$ in the lower panel of Fig.~\ref{susceptibility} to analyze the Curie-Weiss behavior: the dashed line is a linear fit to the data below 70\,K which yields a fluctuating moment of 1.06\,\muB/atom and a very small negative Curie-Weiss temperature. The other samples have larger values since the slope of $\chi^{-1}$ vs $T$ is lower (cf. Fig.~\ref{susceptibility}, lower panel). This fluctuating moment is much larger than the induced magnetic moment of about 0.055\,\muB/atom measured at 7\,T (see Fig.~\ref{magnetization}). This is a common property of itinerant magnets~\cite{Rhodes1963}.
\subsection{Magnetization and Arrott plots}
\begin{figure}[b]
\begin{center}
\includegraphics[width=0.65\columnwidth,angle=-90]{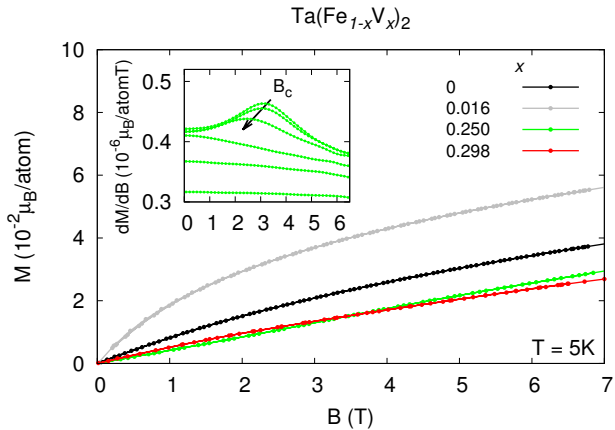}
\end{center}
\caption{Isothermal magnetization $M(B)$ for selected samples with $x$ = 0, 0.016, 0.250 and 0.298. The $M(B)$ curves for $x$ = 0, 0.016 and 0.298 exhibit paramagnetic behavior with a very small induced moment of a few $10^{-2}$\,\muB. The curve for the sample with $x = 0.250$ shows a weak increase of the magnetization at a critical field $B_{\textrm{c}} \approx 3.5$\,T. This is corroborated by the peak in $dM/dB$ vs $B$ plotted in the inset; these measurements were taken at 2, 5, 10, 20, 30 and 50\,K.}
\label{magnetization}
\end{figure}
More evidence for the presence of the SDW state is given by the field dependent magnetization at 5\,K shown in Fig.~\ref{magnetization} for three PM samples and one SDW sample with $x = 0.250$. The isothermal magnetization $M(B)$ of this sample shows a metamagnetic-like increase around $B_{\textrm{c}} \approx 3.5$\,T. This increase is very small, $\Delta M \approx 10^{-3}$\,\muB/atom, and can be considered to be as large as the ordered moment within the SDW phase. The large discrepancy between this value and the effective fluctuating moment extracted from the Curie-Weiss fit, which is larger than 1\,\muB/atom, confirms that the ordered state is a SDW and not AFM from local Fe moments. The transition at $B_{\textrm{c}}$ can be clearly seen in the field derivative $dM/dB$ plotted versus $B$ in the inset of Fig.~\ref{magnetization}. $dM/dB$ shows a distinct peak which shifts to lower fields with increasing temperature, as expected for an antiferromagnet. The $M(B)$ curves for the other samples do not show any feature up to 7\,T as expected since they are PM. The induced moment at 7\,T is in all samples very small, of the order of $10^{-2}$\,\muB/atom. This moment increases with decreasing $x$ and shows its maximum value at $x = 0.016$. This value is almost twice that for the sample with $x = 0.298$.
\begin{figure}[t]
\begin{center}
\includegraphics[width=\columnwidth]{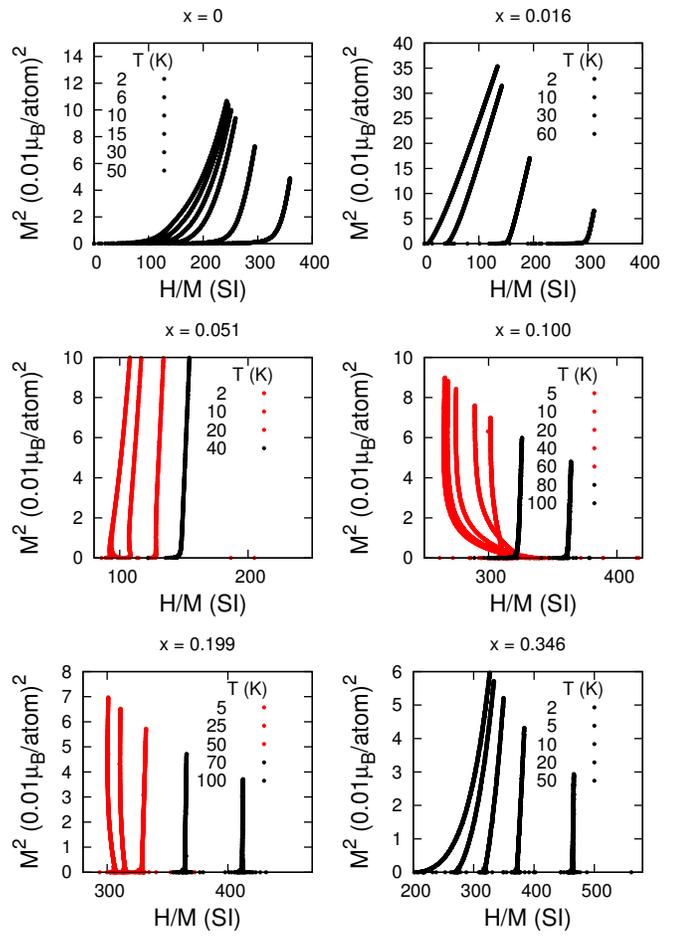}
\end{center}
\caption{Arrott plots for \TFV\ with $H = B/\mu_{\textrm{0}}$. The data in the PM state are plotted with black points whereas those within the SDW state are plotted with red points. This way of plotting emphasizes the opposite curvature between curves in the PM state and those in the SDW state.}
\label{Arrott_plots}
\end{figure}

The magnetic equation of state in nearly FM metals can be expressed by expanding the Ginzburg-Landau free energy $F(B)$~\cite{Moriya1985,Lonzarich1985}:
\begin{equation}
B(M) = \frac{1}{V}\frac{\partial F}{\partial M} = A(T)M(T,B) + bM^{3}(T,B) 
\end{equation}
where
\begin{equation}
A(T) = a + b[3<m_{\parallel}^{2}> + 2<m_{\perp}^{2}>] \sim k_{\textrm{B}}T
\label{A}
\end{equation}
with $<m_{\parallel}^{2}>$ and $<m_{\perp}^{2}>$ are the thermal spin fluctuations longitudinal and transverse to the local magnetization and are direct proportional to temperature. The parameter $a = \chi_{0}^{-1} = \chi^{-1}(T \rightarrow 0)$ is the inverse initial susceptibility and $b$ is the mode-mode coupling parameter. These parameters can be determined by plotting the inverse magnetic uniform susceptibility $B/M$ as a function of $M^{2}$
\begin{equation}\label{arrott0}
\chi^{-1} = \frac{B}{M} = A + bM^{2}
\end{equation} 
which is known as Arrott plot. These plots are usually used to determine the exact Curie temperature in ferromagnets:  In fact, the high-temperature and high-field $M(B)$ curves plotted in this way are straight lines, reflecting the mean field behavior between $\chi$ and $M$. These lines can be extrapolated towards $B/M = 0$ to extract the $M^{2}(0)$ value of the ordered moment. The Curie temperature is then defined at the temperature at which $M^{2}(0)$ starts to become positive. We have made Arrott plots for selected samples in Fig.~\ref{Arrott_plots} by plotting $M^{2}$ vs $H/M$ with $H = B/\mu_{\textrm{0}}$. For these samples we did not observe positive values of $M^{2}(0)$, which implies that all samples are not FM. For other samples close to stoichiometry in Table~\ref{table} we did observe a positive $M^{2}(0)$ below the FM transition temperature. In samples which undergo SDW order, the linear dependence of $M^{2}$ vs $H/M$ changes over an arc with negative slope (cf. red points in Fig.~\ref{Arrott_plots}) for $T <$\,\TN\ and $H < H_{\textrm{c}}$ with $H_{\textrm{c}} = B_{\textrm{c}}/\mu_{\textrm{0}}$ the critical field needed to suppress the SDW order. These plots clearly demonstrate that the samples with $x$ =0, 0.016 and 0.346 are PM down to 2\,K while the samples with $x$ = 0.051, 0.100 and 0.199 shows SDW order below the respective \TN.
\begin{figure}[t]
\begin{center}
\includegraphics[width=0.65\columnwidth, angle=-90]{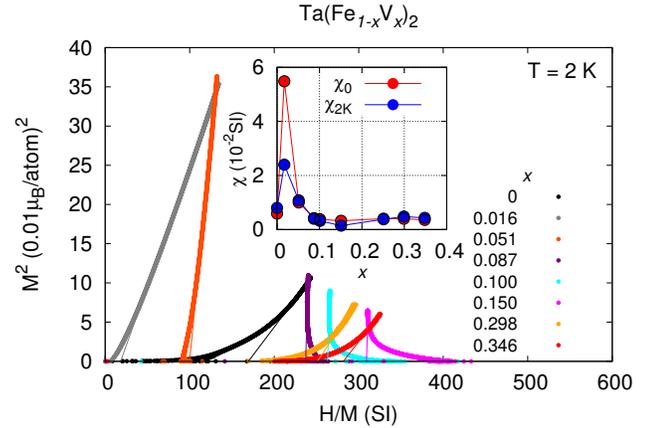}
\end{center}
\caption{Arrott plots for \TFV\ at 2\,K. From this plot we can estimate the mean-field value of the susceptibility $A \approx a = \chi^{-1}_{0}$ from the high-field magnetization according to Eq.~\ref{arrott0}. These values are plotted in the inset as red points. The blue points are the measured susceptibilities at 2\,K, $\chi_{\textrm{2K}}$.}
\label{Arrott_chi0}
\end{figure}

The high-$T$ and high-$B$ mean field behavior very often is not observed at low field and low temperature, in particular in systems with a relevant amount of FM impurities or in systems close to a QCP~\cite{Franz2010}. However, we can extrapolate our measurements from high fields at the lowest measured temperature of 2\,K to obtain the value of the mean field susceptibility $A \approx a = \chi^{-1}_{0}$ and compare it with that directly measured at 2\,K and 1\,T, $\chi_{\textrm{2K}}$. The extrapolation for all samples is shown in Fig.~\ref{Arrott_chi0} and the values for $\chi_{0}$ and $\chi_{\textrm{2K}}$ are plotted in the inset of the same figure as a function of $x$. We first note that $\chi_{0} > \chi_{\textrm{2K}}$ for samples which at 2\,K are in the SDW phase and $\chi_{0} < \chi_{\textrm{2K}}$ for samples in the PM state because of the opposite curvature, as explained before. This is not valid only for the sample with $x = 0.016$ because we have taken the measured susceptibility at 1\,T from Fig.~\ref{susceptibility}. If we would take the susceptibility at much less fields than $\chi_{0} < \chi_{\textrm{2K}}$ for this sample as well. But what we are interested in is the general behavior of these quantities. In fact, both values behave very similar: At high vanadium content a broad maximum is found around $x = 0.3$ at which the SDW order vanishes. There is a minimum at $x \approx 0.15$ where the SDW order has the largest \TN, then both susceptibilities increase steeply for $x < 0.15$ displaying a sharp peak at $x = 0.016$. This behavior indicates that below $x = 0.15$ FM correlations start to grow rapidly. This is the reason for the observed decrease in \TN\ below $x = 0.13$.
\section{Phase diagram and multicriticality}
\begin{figure}[b]
\begin{center}
\includegraphics[width=0.65\columnwidth, angle=-90]{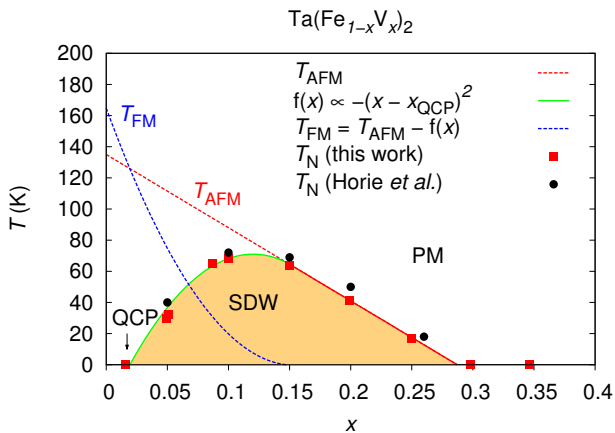}
\end{center}
\caption{Phasediagram of \TFV\ derived from resistivity and susceptibility measurements at 1\,T. Data from Ref.~\citeonline{Horie2010} also also included. The orange area indicates the dome-like SDW phase. The red and blue dashed lines represent the energy scales related to the AFM and FM correlations, respectively.}
\label{phase_diagram}
\end{figure}

From our resistivity (at $B = 0$) and susceptibility (at $B = 1$\,T) measurements we have determined the magnetic phase diagram of \TFV\ which is displayed in Fig.~\ref{phase_diagram}. The red points indicate the N\'eel temperatures observed in our experiments, while the black ones have been extracted from Ref.~\citeonline{Horie2010}, where the susceptibility has also been measured at 1\,T. There is a certain systematic discrepancy between these points, which is difficult to clarify. It could be explained by the fact that our V content has been estimated by chemical analysis while the nominal one is given in Ref.~\citeonline{Horie2010}. However, if there was a systematic difference in $x$, we would expect our \TN\ to be higher on one side and lower on the other side of the dome maximum. The overall behavior is consistent though.

A clear SDW dome emerges in the PM phase of the $x-T$ phase diagram. This dome is clearly not symmetric and presents two QCPs, one at $x \approx 0.02$ and the second one at $x \approx 0.3$. The one at $x \approx 0.3$ seems to be not interesting since the susceptibility and resistivity do not show any particular NFL behavior. On the other hand, the one at $x \approx 0.02$ is very interesting as we will comment below. By looking at high vanadium content, \TN\ follows a linear behavior which suggests that with decreasing $x$ the energy scale $T_{\textrm{AFM}}$ related to the AFM correlations increases linearly. By extrapolating \TN\ between 0.3 and 0.15 towards $x = 0$ (red dashed line in Fig.~\ref{phase_diagram}), we would expect the stoichiometric system \TF\ to order antiferromagnetically at about 140\,K. However, below $x = 0.15$ \TN\ starts to decrease. This means that there is another energy scale, possibly ferromagnetic, which competes with \TN. We can fit the points for \TN\ between 0.02 and 0.15 with a simple function $f(x) \propto -(x-x_{\textrm{QCP}})^{2}$ (green line in Fig.~\ref{phase_diagram}) with $x_{\textrm{QCP}} = 0.02$. Taking now the difference between these two energy scales we can estimate what is the $x$-dependence of the FM energy scale $T_{\textrm{FM}} = T_{\textrm{AFM}} -f(x)$ (blue dashed line in Fig.~\ref{phase_diagram}). The behavior of $T_{\textrm{FM}}$ vs $x$ recalls very well that of both susceptibilities $\chi_{\textrm{0}}$ and $\chi_{\textrm{2K}}$ in the inset of Fig.~\ref{Arrott_chi0} which both increase rapidly below $x = 0.15$. This comparison simply explain the origin of the dome-like shape of the phase diagram. In addition, the results presented in Fig.~\ref{caratioexpfit} suggest that this effect is strictly related to a significant lattice distortion of the crystal structure manifested in the $c/a$ ratio which shows a minimum at $x \approx 0.13$. 
\begin{figure}[t]
\begin{center}
\includegraphics[width=0.65\columnwidth, angle=-90]{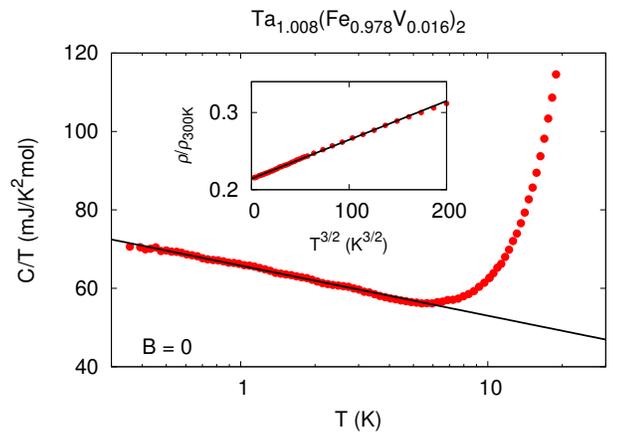}
\end{center}
\caption{Sommerfeld coefficient $C(T)/T$ as a function of the logarithm of the temperature. The inset shows the $T^{3/2}$ behavior of the resistivity at low $T$. It should be noted that the change in resistivity in this temperature range is not very large due to the small RRR.}
\label{multicriticality}
\end{figure}

Turning back to the QCP at $x \approx 0.02$, we can now understand its peculiarity. Both energy scales, $T_{\textrm{FM}}$ and $T_{\textrm{AFM}}$, intersect at this vanadium content leaving a frustrated ground state with high susceptibilities at both wave vectors $q = 0$ and $q = Q_{\mathrm{SDW}}$. This scenario was indeed proposed for \NF\ in Ref.~\citeonline{Moroni2009}. In \NF\ NFL properties were observed at the QCP in form of $\rho \propto T^{3/2}$ and $C/T \propto -\log(T)$. We observe in the sample with $x = 0.016$ located at the QCP the very same behavior as illustrated in Fig.~\ref{multicriticality}. This suggests that the QCP in \TFV\ and the QCP in \NFy\ have the same origin and nature. More interestingly, since more than one energy scale vanishes at the QCP, the nature of this QCP seems to be dual, i.e, it is a quantum multicritical point.
\begin{acknowledgment}
We are indebted to M. Garst, C. Geibel, M. Grosche, D. Kasinathan, D. Rauch, H. Rosner and S. S\"ullow for insightful comments. We would like to thank S. Borisenko for having prepared some of the samples during his school project at the MPI-CPfS.
\end{acknowledgment}
\bibliographystyle{jpsj}
\bibliography{brando_jpsj}

\begin{thebibliography}{10}

\bibitem{Sachdev1999}
S.~Sachdev: {\em Quantum Phase Transitions} (Cambridge University Press,
  Cambridge, 1999).

\bibitem{Stewart2011}
G.~R. Stewart: Rev. Mod. Phys. {\bfseries 83} (2011) 1589.

\bibitem{Broun2008}
D.~M. Broun: Nat. Phys. {\bfseries 4} (2008) 170.

\bibitem{Mathur1998}
N.~D. Mathur, F.~M. Grosche, S.~R. Julian, I.~R. Walker, D.~M. Freye, R.~K.~W.
  Haselwimmer, and G.~G. Lonzarich: Nature {\bfseries 394} (1998) 39.

\bibitem{Yuan2003}
H.~Q. Yuan, F.~M. Grosche, M.~Deppe, C.~Geibel, G.~Sparn, and F.~Steglich:
  Science {\bfseries 302} (2003) 2104.

\bibitem{Baym2004}
G.~Baym and C.~Pethick: {\em Landau Fermi-liquid theory} (Wiley-VCH, 2004).

\bibitem{Yeh2002}
A.~Yeh, Y.-A. Soh, J.~Brooke, G.~Aeppli, T.~F. Rosenbaum, and S.~M. Hayden:
  Nature {\bfseries 419} (2002) 459.

\bibitem{Pfleiderer2007}
C.~Pfleiderer, P.~Böni, T.~Keller, U.~K. Rößler, and A.~Rosch: Science
  {\bfseries 316} (2007) 1871.

\bibitem{Niklowitz2005}
P.~G. Niklowitz, F.~Beckers, G.~G. Lonzarich, G.~Knebel, B.~Salce,
  J.~Thomasson, N.~Bernhoeft, D.~Braithwaite, and J.~Flouquet: Phys. Rev. B
  {\bfseries 72} (2005) 024424.

\bibitem{Smith2008}
R.~P. Smith, M.~Sutherland, G.~G. Lonzarich, S.~S. Saxena, N.~Kimura,
  S.~Takashima, M.~Nohara, and H.~Takagi: Nature {\bfseries 455} (2008) 1220.

\bibitem{Grigera2004}
S.~A. Grigera, P.~Gegenwart, R.~A. Borzi, F.~Weickert, A.~J. Schofield, R.~S.
  Perry, T.~Tayama, T.~Sakakibara, Y.~Maeno, A.~G. Green, and A.~P. Mackenzie:
  Science {\bfseries 306} (2004) 1154.

\bibitem{Yamada1988}
Y.~Yamada and A.~Sakata: J. Phys. Soc. Jpn. {\bfseries 57} (1988) 46.

\bibitem{Crook1995}
M.~R. Crook: {Ph.D.} thesis, University of Reading (1995).

\bibitem{Yamada1990}
Y.~Yamada, H.~Nakamura, Y.~Kitaoka, K.~Asayama, K.~Koga, A.~Sakata, and
  T.~Murakami: J. Phys. Soc. Jpn. {\bfseries 59} (1990) 2976.

\bibitem{Horie2010}
Y.~Horie, S.~Kawashima, Y.~Yamada, G.~Obara, and T.~Nakamura: Journal of
  Physics: Conference Series {\bfseries 200} (2010) 032078.

\bibitem{Moroni2009}
D.~Moroni-Klementowicz, M.~Brando, C.~Albrecht, W.~J. Duncan, F.~M. Grosche,
  D.~Gr\"uner, and G.~Kreiner: Phys. Rev. B {\bfseries 79} (2009) 224410.

\bibitem{Duncan2010}
W.~J. Duncan, O.~P. Welzel, D.~Moroni-Klementowicz, C.~Albrecht, P.~G.
  Niklowitz, D.~Gr{\"u}ner, M.~Brando, A.~Neubauer, C.~Pfleiderer, N.~Kikugawa,
  A.~P. Mackenzie, and F.~M. Grosche: physica status solidi (b) {\bfseries 247}
  (2010) 544.

\bibitem{Rauch2015}
D.~Rauch, M.~Kraken, F.~J. Litterst, S.~S\"ullow, H.~Luetkens, M.~Brando,
  T.~F\"orster, J.~Sichelschmidt, A.~Neubauer, C.~Pfleiderer, W.~J. Duncan, and
  F.~M. Grosche: Phys. Rev. B {\bfseries 91} (2015) 174404.

\bibitem{Brando2008}
M.~Brando, W.~J. Duncan, D.~Moroni-Klementowicz, C.~Albrecht, D.~Gr\"uner,
  R.~Ballou, and F.~M. Grosche: Phys. Rev. Lett. {\bfseries 101} (2008) 026401.

\bibitem{Stoe}
{\em Stoe WinXpow software package} (STOE and CIE GmbH, 2007).

\bibitem{Jana2006}
V.~Petricek, M.~Dusek, and L.~Palatinus: {\em The crystallographic computing
  system} (Institut of Physics, Praha, Czech Republic, 2006).

\bibitem{Guzei1970}
L.~S. Guzei, E.~M. Sokolovs, I.~G. Sokolova, G.~N. Ronami, and S.~M. Kuznetso:
  Vestnik Moskovskogo Universiteta Seriya 2 Khimiya {\bfseries 11} (1970) 696.

\bibitem{Kuo1953}
K.~Kuo: Act. Met. {\bfseries 1} (1953) 720.

\bibitem{Pauling2002}
{\em Pauling File} (Binaries Edition, Vers. 1.0, 2002).

\bibitem{Kerkau2012}
A.~Kerkau: {Ph.D.} thesis, University of Dresden (2013).

\bibitem{Pauling1927}
L.~Pauling: J. Am. Chem. Soc. {\bfseries 49} (1927) 765.

\bibitem{Gruener1994}
G.~Gr\"uner: Rev. Mod. Phys. {\bfseries 66} (1994) 1.

\bibitem{Koepernik1999}
K.~Koepernik and H.~Eschrig: Phys. Rev. B {\bfseries 59} (1999) 1743.

\bibitem{Perdew1992}
J.~P. Perdew and Y.~Wang: Phys. Rev. B {\bfseries 45} (1992) 13244.

\bibitem{Diop2015}
L.~V.~B. Diop, D.~Benea, S.~Mankovsky, and O.~Isnard: Journal of Alloys and
  Compounds {\bfseries 643} (2015) 239.

\bibitem{Moriya1985}
T.~Moriya: {\em Spin Fluctuations in Itinerant Electron Magnetism} (Springer,
  Berlin, 1985).

\bibitem{Rhodes1963}
E.~P.~W. P.~Rhodes: Proceedings of the Royal Society of London. Series A,
  Mathematical and Physical Sciences {\bfseries 273} (1963) 247.

\bibitem{Lonzarich1985}
G.~G. Lonzarich and L.~Taillefer: J. Phys. C {\bfseries 18} (1985) 4339.

\bibitem{Franz2010}
C.~Franz, C.~Pfleiderer, A.~Neubauer, M.~Schulz, B.~Pedersen, and P.~B\"oni: J.
  Phys.: Conf. Series {\bfseries 200} (2010) 012036.

\end{thebibliography}
\end{document}